%
\catcode`@=11 
%
%

\font\fourteenrm=cmr10 scaled\magstep2
\font\twelverm=cmr10 scaled\magstep1
\font\ninerm=cmr9            \font\sixrm=cmr6

\font\fourteenbf=cmbx10 scaled\magstep2
\font\twelvebf=cmbx10 scaled\magstep1
\font\ninebf=cmbx9            \font\sixbf=cmbx6
\font\seventeeni=cmmi10 scaled\magstep3     \skewchar\seventeeni='177
\font\fourteeni=cmmi10 scaled\magstep2      \skewchar\fourteeni='177
\font\twelvei=cmmi10 scaled\magstep1        \skewchar\twelvei='177
\font\ninei=cmmi9                           \skewchar\ninei='177
\font\sixi=cmmi6                            \skewchar\sixi='177
\font\seventeensy=cmsy10 scaled\magstep3    \skewchar\seventeensy='60
\font\fourteensy=cmsy10 scaled\magstep2     \skewchar\fourteensy='60
\font\twelvesy=cmsy10 scaled\magstep1       \skewchar\twelvesy='60
\font\ninesy=cmsy9                          \skewchar\ninesy='60
\font\sixsy=cmsy6                           \skewchar\sixsy='60

\font\fourteenex=cmex10 scaled\magstep2
\font\twelveex=cmex10 scaled\magstep1

\font\fourteensl=cmsl10 scaled\magstep2                                         
\font\twelvesl=cmsl10 scaled\magstep1                                           
\font\ninesl=cmsl9                                                              

\font\fourteenit=cmti10 scaled\magstep2                                         
\font\twelveit=cmti10 scaled\magstep1                                           
\font\twelvett=cmtt10 scaled\magstep1                                           
\font\twelvecp=cmcsc10 scaled\magstep1                                          
\font\tencp=cmcsc10                                                             
\newfam\cpfam                                                                   
%
%
\newcount\f@ntkey            \f@ntkey=0                                         
\def\samef@nt{\relax \ifcase\f@ntkey \rm \or\oldstyle \or\or                    
         \or\it \or\sl \or\bf \or\tt \or\caps \fi }                             
\def\fourteenpoint{\relax                                                       
    \textfont0=\fourteenrm          \scriptfont0=\tenrm                         
    \scriptscriptfont0=\sevenrm                                                 
     \def\rm{\fam0 \fourteenrm \f@ntkey=0 }\relax                               
    \textfont1=\fourteeni           \scriptfont1=\teni                          
    \scriptscriptfont1=\seveni                                                  
     \def\oldstyle{\fam1 \fourteeni\f@ntkey=1 }\relax                           
    \textfont2=\fourteensy          \scriptfont2=\tensy                         
    \scriptscriptfont2=\sevensy                                                 
    \textfont3=\fourteenex     \scriptfont3=\fourteenex                         
    \scriptscriptfont3=\fourteenex                                              
    \def\it{\fam\itfam \fourteenit\f@ntkey=4 }\textfont\itfam=\fourteenit       
    \def\sl{\fam\slfam \fourteensl\f@ntkey=5 }\textfont\slfam=\fourteensl       
    \scriptfont\slfam=\tensl                                                    
    \def\bf{\fam\bffam \fourteenbf\f@ntkey=6 }\textfont\bffam=\fourteenbf       
    \scriptfont\bffam=\tenbf     \scriptscriptfont\bffam=\sevenbf               
    \def\tt{\fam\ttfam \twelvett \f@ntkey=7 }\textfont\ttfam=\twelvett          
    \h@big=11.9\p@{} \h@Big=16.1\p@{} \h@bigg=20.3\p@{} \h@Bigg=24.5\p@{}       
    \def\caps{\fam\cpfam \twelvecp \f@ntkey=8 }\textfont\cpfam=\twelvecp        
    \setbox\strutbox=\hbox{\vrule height 12pt depth 5pt width\z@}               
    \samef@nt}                                                                  
\def\twelvepoint{\relax                                                         
    \textfont0=\twelverm          \scriptfont0=\ninerm                          
    \scriptscriptfont0=\sixrm                                                   
     \def\rm{\fam0 \twelverm \f@ntkey=0 }\relax                                 
    \textfont1=\twelvei           \scriptfont1=\ninei                           
    \scriptscriptfont1=\sixi                                                    
     \def\oldstyle{\fam1 \twelvei\f@ntkey=1 }\relax                             
    \textfont2=\twelvesy          \scriptfont2=\ninesy                          
    \scriptscriptfont2=\sixsy                                                   
    \textfont3=\twelveex          \scriptfont3=\twelveex                        
    \scriptscriptfont3=\twelveex                                                
    \def\it{\fam\itfam \twelveit \f@ntkey=4 }\textfont\itfam=\twelveit          
    \def\sl{\fam\slfam \twelvesl \f@ntkey=5 }\textfont\slfam=\twelvesl          
    \scriptfont\slfam=\ninesl                                                   
    \def\bf{\fam\bffam \twelvebf \f@ntkey=6 }\textfont\bffam=\twelvebf          
    \scriptfont\bffam=\ninebf     \scriptscriptfont\bffam=\sixbf                
    \def\tt{\fam\ttfam \twelvett \f@ntkey=7 }\textfont\ttfam=\twelvett          
    \h@big=10.2\p@{}                                                            
    \h@Big=13.8\p@{}                                                            
    \h@bigg=17.4\p@{}                                                           
    \h@Bigg=21.0\p@{}                                                           
    \def\caps{\fam\cpfam \twelvecp \f@ntkey=8 }\textfont\cpfam=\twelvecp        
    \setbox\strutbox=\hbox{\vrule height 10pt depth 4pt width\z@}               
    \samef@nt}                                                                  
\def\tenpoint{\relax                                                            
    \textfont0=\tenrm          \scriptfont0=\sevenrm                            
    \scriptscriptfont0=\fiverm                                                  
    \def\rm{\fam0 \tenrm \f@ntkey=0 }\relax                                     
    \textfont1=\teni           \scriptfont1=\seveni                             
    \scriptscriptfont1=\fivei                                                   
    \def\oldstyle{\fam1 \teni \f@ntkey=1 }\relax                                
    \textfont2=\tensy          \scriptfont2=\sevensy                            
    \scriptscriptfont2=\fivesy                                                  
    \textfont3=\tenex          \scriptfont3=\tenex                              
    \scriptscriptfont3=\tenex                                                   
    \def\it{\fam\itfam \tenit \f@ntkey=4 }\textfont\itfam=\tenit                
    \def\sl{\fam\slfam \tensl \f@ntkey=5 }\textfont\slfam=\tensl                
    \def\bf{\fam\bffam \tenbf \f@ntkey=6 }\textfont\bffam=\tenbf                
    \scriptfont\bffam=\sevenbf     \scriptscriptfont\bffam=\fivebf              
    \def\tt{\fam\ttfam \tentt \f@ntkey=7 }\textfont\ttfam=\tentt                
    \def\caps{\fam\cpfam \tencp \f@ntkey=8 }\textfont\cpfam=\tencp              
    \setbox\strutbox=\hbox{\vrule height 8.5pt depth 3.5pt width\z@}            
    \samef@nt}                                                                  
%
%
%
%
\newdimen\h@big  \h@big=8.5\p@                                                  
\newdimen\h@Big  \h@Big=11.5\p@                                                 
\newdimen\h@bigg  \h@bigg=14.5\p@                                               
\newdimen\h@Bigg  \h@Bigg=17.5\p@                                               
\def\big#1{{\hbox{$\left#1\vbox to\h@big{}\right.\n@space$}}}                   
\def\Big#1{{\hbox{$\left#1\vbox to\h@Big{}\right.\n@space$}}}                   
\def\bigg#1{{\hbox{$\left#1\vbox to\h@bigg{}\right.\n@space$}}}                 
\def\Bigg#1{{\hbox{$\left#1\vbox to\h@Bigg{}\right.\n@space$}}}                 
%
%
%
\normalbaselineskip = 20pt plus 0.2pt minus 0.1pt                               
\normallineskip = 1.5pt plus 0.1pt minus 0.1pt                                  
\normallineskiplimit = 1.5pt                                                    
\newskip\normaldisplayskip                                                      
\normaldisplayskip = 20pt plus 5pt minus 10pt                                   
\newskip\normaldispshortskip                                                    
\normaldispshortskip = 6pt plus 5pt                                             
\newskip\normalparskip                                                          
\normalparskip = 6pt plus 2pt minus 1pt                                         
\newskip\skipregister                                                           
\skipregister = 5pt plus 2pt minus 1.5pt                                        
\newif\ifsingl@    \newif\ifdoubl@                                              
\newif\iftwelv@    \twelv@true                                                  
\def\singlespace{\singl@true\doubl@false\spaces@t}                              
\def\doublespace{\singl@false\doubl@true\spaces@t}                              
\def\normalspace{\singl@false\doubl@false\spaces@t}                             
\def\Tenpoint{\tenpoint\twelv@false\spaces@t}                                   
\def\Twelvepoint{\twelvepoint\twelv@true\spaces@t}                              
\def\spaces@t{\relax                                                            
      \iftwelv@ \ifsingl@\subspaces@t3:4;\else\subspaces@t1:1;\fi               
       \else \ifsingl@\subspaces@t3:5;\else\subspaces@t4:5;\fi \fi              
      \ifdoubl@ \multiply\baselineskip by 5                                     
         \divide\baselineskip by 4 \fi }                                        
\def\subspaces@t#1:#2;{                                                         
      \baselineskip = \normalbaselineskip                                       
      \multiply\baselineskip by #1 \divide\baselineskip by #2                   
      \lineskip = \normallineskip                                               
      \multiply\lineskip by #1 \divide\lineskip by #2                           
      \lineskiplimit = \normallineskiplimit                                     
      \multiply\lineskiplimit by #1 \divide\lineskiplimit by #2                 
      \parskip = \normalparskip                                                 
      \multiply\parskip by #1 \divide\parskip by #2                             
      \abovedisplayskip = \normaldisplayskip                                    
      \multiply\abovedisplayskip by #1 \divide\abovedisplayskip by #2           
      \belowdisplayskip = \abovedisplayskip                                     
      \abovedisplayshortskip = \normaldispshortskip                             
      \multiply\abovedisplayshortskip by #1                                     
        \divide\abovedisplayshortskip by #2                                     
      \belowdisplayshortskip = \abovedisplayshortskip                           
      \advance\belowdisplayshortskip by \belowdisplayskip                       
      \divide\belowdisplayshortskip by 2                                        
      \smallskipamount = \skipregister                                          
      \multiply\smallskipamount by #1 \divide\smallskipamount by #2             
      \medskipamount = \smallskipamount \multiply\medskipamount by 2            
      \bigskipamount = \smallskipamount \multiply\bigskipamount by 4 }          
\def\normalbaselines{ \baselineskip=\normalbaselineskip                         
   \lineskip=\normallineskip \lineskiplimit=\normallineskip                     
   \iftwelv@\else \multiply\baselineskip by 4 \divide\baselineskip by 5         
     \multiply\lineskiplimit by 4 \divide\lineskiplimit by 5                    
     \multiply\lineskip by 4 \divide\lineskip by 5 \fi }                        
\Twelvepoint  
\interlinepenalty=50                                                            
\interfootnotelinepenalty=5000                                                  
\predisplaypenalty=9000                                                         
\postdisplaypenalty=500                                                         
\hfuzz=1pt                                                                      
\vfuzz=0.2pt                                                                    
%
%
%
\def\pagecontents{                                                              
   \ifvoid\topins\else\unvbox\topins\vskip\skip\topins\fi                       
   \dimen@ = \dp255 \unvbox255                                                  
   \ifvoid\footins\else\vskip\skip\footins\footrule\unvbox\footins\fi           
   \ifr@ggedbottom \kern-\dimen@ \vfil \fi }                                    
\def\makeheadline{\vbox to 0pt{ \skip@=\topskip                                 
      \advance\skip@ by -12pt \advance\skip@ by -2\normalbaselineskip           
      \vskip\skip@ \line{\vbox to 12pt{}\the\headline} \vss                     
      }\nointerlineskip}                                                        
\def\makefootline{\baselineskip = 1.5\normalbaselineskip                        
                 \line{\the\footline}}                                          
\newif\iffrontpage                                                              
\newif\ifletterstyle                                                            
\newif\ifp@genum                                                                
\def\nopagenumbers{\p@genumfalse}                                               
\def\pagenumbers{\p@genumtrue}                                                  
\pagenumbers                                                                    
\newtoks\paperheadline                                                          
\newtoks\letterheadline                                                         
\newtoks\letterfrontheadline                                                    
\newtoks\lettermainheadline                                                     
\newtoks\paperfootline                                                          
\newtoks\letterfootline                                                         
\newtoks\date                                                                   
\footline={\ifletterstyle\the\letterfootline\else\the\paperfootline\fi}         
\paperfootline={\hss\iffrontpage\else\ifp@genum\tenrm\folio\hss\fi\fi}          
\letterfootline={\hfil}                                                         
\headline={\ifletterstyle\the\letterheadline\else\the\paperheadline\fi}         
\paperheadline={\hfil}                                                          
\letterheadline{\iffrontpage\the\letterfrontheadline                            
     \else\the\lettermainheadline\fi}                                           
\lettermainheadline={\rm\ifp@genum page \ \folio\fi\hfil\the\date}              
\def\monthname{\relax\ifcase\month 0/\or January\or February\or                 
   March\or April\or May\or June\or July\or August\or September\or              
   October\or November\or December\else\number\month/\fi}                       
\date={\monthname\ \number\day, \number\year}                                   
\countdef\pagenumber=1  \pagenumber=1                                           
\def\advancepageno{\global\advance\pageno by 1                                  
   \ifnum\pagenumber<0 \global\advance\pagenumber by -1                         
    \else\global\advance\pagenumber by 1 \fi \global\frontpagefalse }           
\def\folio{\ifnum\pagenumber<0 \romannumeral-\pagenumber                        
           \else \number\pagenumber \fi }                                       
\def\footrule{\dimen@=\prevdepth\nointerlineskip                                
   \vbox to 0pt{\vskip -0.25\baselineskip \hrule width 0.35\hsize \vss}         
   \prevdepth=\dimen@ }                                                         
\newtoks\foottokens                                                             
\foottokens={\Tenpoint\singlespace}                                             
\newdimen\footindent                                                            
\footindent=24pt                                                                
\def\vfootnote#1{\insert\footins\bgroup  \the\foottokens                        
   \interlinepenalty=\interfootnotelinepenalty \floatingpenalty=20000           
   \splittopskip=\ht\strutbox \boxmaxdepth=\dp\strutbox                         
   \leftskip=\footindent \rightskip=\z@skip                                     
   \parindent=0.5\footindent \parfillskip=0pt plus 1fil                         
   \spaceskip=\z@skip \xspaceskip=\z@skip                                       
   \Textindent{$ #1 $}\footstrut\futurelet\next\fo@t}                           
\def\Textindent#1{\noindent\llap{#1\enspace}\ignorespaces}                      
\def\footnote#1{\attach{#1}\vfootnote{#1}}

\let\footsymbol=\star                                                           
\newcount\lastf@@t           \lastf@@t=-1                                       
\newcount\footsymbolcount    \footsymbolcount=0                                 
\newif\ifPhysRev                                                                
\def\footsymbolgen{\relax \ifPhysRev \iffrontpage \NPsymbolgen\else             
      \PRsymbolgen\fi \else \NPsymbolgen\fi                                     
   \global\lastf@@t=\pageno \footsymbol }                                       
\def\NPsymbolgen{\ifnum\footsymbolcount<0 \global\footsymbolcount=0\fi          
   {\iffrontpage \else \advance\lastf@@t by 1 \fi                               
    \ifnum\lastf@@t<\pageno \global\footsymbolcount=0                           
     \else \global\advance\footsymbolcount by 1 \fi }                           
   \ifcase\footsymbolcount \fd@f\star\or \fd@f\dagger\or \fd@f\ast\or           
    \fd@f\ddagger\or \fd@f\natural\or \fd@f\diamond\or \fd@f\bullet\or          
    \fd@f\nabla\else \fd@f\dagger\global\footsymbolcount=0 \fi }                
\def\fd@f#1{\xdef\footsymbol{#1}}                                               
\def\PRsymbolgen{\ifnum\footsymbolcount>0 \global\footsymbolcount=0\fi          
      \global\advance\footsymbolcount by -1                                     
      \xdef\footsymbol{\sharp\number-\footsymbolcount} }                        
\def\space@ver#1{\let\@sf=\empty \ifmmode #1\else \ifhmode                      
   \edef\@sf{\spacefactor=\the\spacefactor}\unskip${}#1$\relax\fi\fi}           
\def\attach#1{\space@ver{\strut^{\mkern 2mu #1} }\@sf\ }                        
%
%
%
\newcount\chapternumber      \chapternumber=0                                   
\newcount\sectionnumber      \sectionnumber=0                                   
\newcount\equanumber         \equanumber=0                                      
\let\chapterlabel=0                                                             
\newtoks\chapterstyle        \chapterstyle={\Number}                            
\newskip\chapterskip         \chapterskip=\bigskipamount                        
\newskip\sectionskip         \sectionskip=\medskipamount                        
\newskip\headskip            \headskip=8pt plus 3pt minus 3pt                   
\newdimen\chapterminspace    \chapterminspace=15pc                              
\newdimen\sectionminspace    \sectionminspace=10pc                              
\newdimen\referenceminspace  \referenceminspace=25pc                            
\def\chapterreset{\global\advance\chapternumber by 1                            
   \ifnum\equanumber<0 \else\global\equanumber=0\fi                             
   \sectionnumber=0 \makel@bel}                                                 
\def\makel@bel{\xdef\chapterlabel{%
\the\chapterstyle{\the\chapternumber}.}}                                        
\def\sectionlabel{\number\sectionnumber \quad }                                 
\def\alphabetic#1{\count255='140 \advance\count255 by #1\char\count255}         
\def\Alphabetic#1{\count255='100 \advance\count255 by #1\char\count255}         
\def\Roman#1{\uppercase\expandafter{\romannumeral #1}}                          
\def\roman#1{\romannumeral #1}                                                  
\def\Number#1{\number #1}                                                       
\def\unnumberedchapters{\let\makel@bel=\relax \let\chapterlabel=\relax          
\let\sectionlabel=\relax \equanumber=-1 }                                       
\def\titlestyle#1{\par\begingroup \interlinepenalty=9999                        
     \leftskip=0.02\hsize plus 0.23\hsize minus 0.02\hsize                      
     \rightskip=\leftskip \parfillskip=0pt                                      
     \hyphenpenalty=9000 \exhyphenpenalty=9000                                  
     \tolerance=9999 \pretolerance=9000                                         
     \spaceskip=0.333em \xspaceskip=0.5em                                       
     \iftwelv@\fourteenpoint\else\twelvepoint\fi                                
   \noindent #1\par\endgroup }                                                  
\def\spacecheck#1{\dimen@=\pagegoal\advance\dimen@ by -\pagetotal               
   \ifdim\dimen@<#1 \ifdim\dimen@>0pt \vfil\break \fi\fi}                       
\def\chapter#1{\par \penalty-300 \vskip\chapterskip                             
   \spacecheck\chapterminspace                                                  
   \chapterreset \titlestyle{\chapterlabel \ #1}                                
   \nobreak\vskip\headskip \penalty 30000                                       
   \wlog{\string\chapter\ \chapterlabel} }                                      

\def\section#1{\par \ifnum\the\lastpenalty=30000\else                           
   \penalty-200\vskip\sectionskip \spacecheck\sectionminspace\fi                
   \wlog{\string\section\ \chapterlabel \the\sectionnumber}                     
   \global\advance\sectionnumber by 1  \noindent                                
   {\caps\enspace\chapterlabel \sectionlabel #1}\par                            
   \nobreak\vskip\headskip \penalty 30000 }                                     
\def\subsection#1{\par                                                          
   \ifnum\the\lastpenalty=30000\else \penalty-100\smallskip \fi                 
   \noindent\undertext{#1}\enspace \vadjust{\penalty5000}}                      

\def\undertext#1{\vtop{\hbox{#1}\kern 1pt \hrule}}                              
\def\APPENDIX#1#2{\par\penalty-300\vskip\chapterskip                            
   \spacecheck\chapterminspace \chapterreset \xdef\chapterlabel{#1}             
   \titlestyle{APPENDIX #2} \nobreak\vskip\headskip \penalty 30000              
   \wlog{\string\Appendix\ \chapterlabel} }                                     
\def\Appendix#1{\APPENDIX{#1}{#1}}                                              
\def\appendix{\APPENDIX{A}{}}                                                   
%
%
%
\def\eqname#1{\relax \ifnum\equanumber<0                                        
     \xdef#1{{\rm(\number-\equanumber)}}\global\advance\equanumber by -1        
    \else \global\advance\equanumber by 1                                       
      \xdef#1{{\rm(\chapterlabel \number\equanumber)}} \fi}                     
\def\eqinsert#1{\noalign{\dimen@=\prevdepth \nointerlineskip                    
   \setbox0=\hbox to\displaywidth{\hfil #1}                                     
   \vbox to 0pt{\vss\hbox{$\!\box0\!$}\kern-0.5\baselineskip}                   
   \prevdepth=\dimen@}}                                                         
%

%
                                                  
%
                                        
%
%
\def\GENITEM#1;#2{\par \hangafter=0 \hangindent=#1                              
    \Textindent{$ #2 $}\ignorespaces}                                           
\outer\def\newitem#1=#2;{\gdef#1{\GENITEM #2;}}                                 
\newdimen\itemsize                \itemsize=30pt                                
\newitem\item=1\itemsize;                                                       
\newitem\sitem=1.75\itemsize;                                
\newitem\ssitem=2.5\itemsize;                             
\outer\def\newlist#1=#2&#3&#4;{\toks0={#2}\toks1={#3}%
   \count255=\escapechar \escapechar=-1                                         
   \alloc@0\list\countdef\insc@unt\listcount     \listcount=0                   
   \edef#1{\par                                                                 
      \countdef\listcount=\the\allocationnumber                                 
      \advance\listcount by 1                                                   
      \hangafter=0 \hangindent=#4                                               
      \Textindent{\the\toks0{\listcount}\the\toks1}}                            
   \expandafter\expandafter\expandafter                                         
    \edef\c@t#1{begin}{\par                                                     
      \countdef\listcount=\the\allocationnumber \listcount=1                    
      \hangafter=0 \hangindent=#4                                               
      \Textindent{\the\toks0{\listcount}\the\toks1}}                            
   \expandafter\expandafter\expandafter                                         
    \edef\c@t#1{con}{\par \hangafter=0 \hangindent=#4 \noindent}                
   \escapechar=\count255}                                                       
\def\c@t#1#2{\csname\string#1#2\endcsname}                                      
\newlist\point=\Number&.&1.0\itemsize;                                          
\newlist\subpoint=(\alphabetic&)&1.75\itemsize;                                 
\newlist\subsubpoint=(\roman&)&2.5\itemsize;                                    
%

%
%
%
\newcount\referencecount     \referencecount=0                                  
\newif\ifreferenceopen       \newwrite\referencewrite                           
\newtoks\rw@toks                                                                
\def\NPrefmark#1{\attach{\scriptscriptstyle [ #1 ] }}
\let\PRrefmark\attach
\def\refmark#1{\relax\ifPhysRev\PRrefmark{#1}\else\NPrefmark{#1}\fi}            
\def\refend{\refmark{\number\referencecount}}                                   
\newcount\lastrefsbegincount \lastrefsbegincount=0                              
\def\refsend{\refmark{\count255=\referencecount                                 
   \advance\count255 by-\lastrefsbegincount                                     
   \ifcase\count255 \number\referencecount                                      
   \or \number\lastrefsbegincount,\number\referencecount                        
   \else \number\lastrefsbegincount-\number\referencecount \fi}}                
\def\ref#1{\REF\?{#1}\refend}

\def\refch@ck{\chardef\rw@write=\referencewrite                                 
   \ifreferenceopen \else \referenceopentrue                                    
   \immediate\openout\referencewrite=referenc.aux \fi}                     
\def\rw@begin#1\splitout{\rw@toks={#1}\relax                                    
   \immediate\write\rw@write{\the\rw@toks}\futurelet\n@xt\rw@next}              
\def\rw@next{\ifx\n@xt\rw@end \let\n@xt=\relax                                  
      \else \let\n@xt=\rw@begin \fi \n@xt}                                      
\let\rw@end=\relax                                                              
\let\splitout=\relax                                                            
\newdimen\refindent     \refindent=30pt                                         
\def\refitem#1{\par \hangafter=0 \hangindent=\refindent \Textindent{#1}}        
\def\REF#1#2{\space@ver{}\refch@ck                                              
   \global\advance\referencecount by 1 \xdef#1{\the\referencecount}%
   \immediate\write\referencewrite{\noexpand\refitem{#1.}}%
   \rw@begin #2\splitout\rw@end \@sf}                                           
\newcount\figurecount     \figurecount=0                                        
\newif\iffigureopen       \newwrite\figurewrite                                 
\def\figch@ck{\chardef\rw@write=\figurewrite \iffigureopen\else                 
   \immediate\openout\figurewrite=figures.aux                              
   \figureopentrue\fi}                                                          
\def\FIG#1#2{\figch@ck                                                          
   \global\advance\figurecount by 1 \xdef#1{\the\figurecount}%
   \immediate\write\figurewrite{\noexpand\refitem{#1.}}%
   \rw@begin #2\splitout\rw@end}                                                
%
                                                                            
%

%
%
%
\def\ugsfig#1 #2 #3 #4 {\midinsert
\centerline{
\special {insert #1}
\hbox spread #2 {}}
\vskip #3
\medskip
\centerline{Fig. #4}
\endinsert}
\newcount\tablecount     \tablecount=0                                          
\newif\iftableopen       \newwrite\tablewrite                                   
\def\tabch@ck{\chardef\rw@write=\tablewrite \iftableopen\else                   
   \immediate\openout\tablewrite=tables.aux                                
   \tableopentrue\fi}                                                           
\gdef\TABLE#1#2{\tabch@ck                                                       
   \global\advance\tablecount by 1 \xdef#1{\the\tablecount}%
   \immediate\write\tablewrite{\noexpand\refitem{#1.}}%
   \rw@begin #2\splitout\rw@end }                                               
                                            
%
                                                                            
%
%
%
\def\masterreset{\global\pagenumber=1 \global\chapternumber=0                   
   \global\equanumber=0 \global\sectionnumber=0                                 
   \global\referencecount=0 \global\figurecount=0 \global\tablecount=0 }        
\def\FRONTPAGE{\ifvoid255\else\vfill\penalty-2000\fi                            
      \masterreset\global\frontpagetrue                                         
      \global\lastf@@t=0 \global\footsymbolcount=0}                             

\def\paperstyle{\letterstylefalse\normalspace\papersize}                        
\def\letterstyle{\letterstyletrue\singlespace\lettersize}                       
\def\papersize{\hsize=35pc\vsize=50pc\hoffset=8pc\voffset=8pc                   
               \skip\footins=\bigskipamount}                                    
\def\lettersize{\hsize=6.5in\vsize=8.5in\hoffset=1in\voffset=1in
   \skip\footins=\smallskipamount \multiply\skip\footins by 3 }                 
\paperstyle   
%
%
\def\MEMO{\letterstyle\FRONTPAGE \letterfrontheadline={\hfil}                   
    \line{\quad\fourteenrm MEMORANDUM\hfil\twelverm\the\date\quad}         
    \medskip \memod@f}                                                          

\def\memit@m#1{\smallskip \hangafter=0 \hangindent=1in                          
      \Textindent{\caps #1}}                                                    
\def\memod@f{\xdef\to{\memit@m{To:}}\xdef\from{\memit@m{From:}}%
     \xdef\topic{\memit@m{Topic:}}\xdef\subject{\memit@m{Subject:}}%
     \xdef\rule{\bigskip\hrule height 1pt\bigskip}}                             
\memod@f                                                                        
%

%
%
%
%
\font\smallheadfont=cmss10 at 10truept
\font\largeheadfont=cmr10 at 20.74truept 
\newdimen\longinden \longinden=0.5 truein	
\newdimen\leftmarg \leftmarg=\hoffset \advance \leftmarg by -.5in
\newdimen\tmp
\newbox\headbox
\newbox\tmpbox
\setbox\headbox=
   \vbox {\offinterlineskip	
      \hbox to7.5in{}
      \vbox {
        \hbox {\hskip\longinden {\largeheadfont Northeastern University}}}
      \vskip 3truept
      \hrule height2pt \vskip1pt \hrule 
      \vskip 5truept
      \vbox {
        \hbox {\hskip\longinden {\smallheadfont 360 Huntington Avenue, Boston,
      Massachusetts, 02115}}}
      \vskip 9truept
      \hrule
      \vskip 5truept
      \vbox {
        \hbox {\hskip\longinden {\smallheadfont Department of Physics}}
        \vskip 3.5pt \hbox {\hskip\longinden 
{\smallheadfont J. B. Sokoloff
\hskip3em\relax  Tel. (617) 373-2931  \hskip3em\relax 
Internet: 3630jbs@neu.edu \hskip3em\relax FAX (617) 373-2943}}}
      \vskip 20truept
      \hrule \vskip1pt \hrule height2pt
    }
\tmp=\ht\headbox
\advance \tmp by -\voffset	
\advance \tmp by 0.5truein
\setbox\tmpbox=\vbox to \tmp{\vss \box\headbox}
\def\depthead{\vskip-\lastskip
    \moveleft\leftmarg\box\tmpbox\bigskip}
\newskip\lettertopfil                                                           
\lettertopfil = 0pt plus 1.5in minus 0pt                                        
\newskip\letterbottomfil                                                        
\letterbottomfil = 0pt plus 2.3in minus 0pt                                     
\newskip\spskip \setbox0\hbox{\ } \spskip=-1\wd0                                
\def\addressee#1{\medskip\rightline{\the\date\hskip 30pt} \bigskip              
   \vskip\lettertopfil                                                          
   \ialign to\hsize{\strut ##\hfil\tabskip 0pt plus \hsize \cr #1\crcr}         
   \medskip\noindent\hskip\spskip}                                              
\newskip\signatureskip       \signatureskip=40pt                                
\def\signed#1{\par \penalty 9000 \bigskip \dt@pfalse                            
  \everycr={\noalign{\ifdt@p\vskip\signatureskip\global\dt@pfalse\fi}}          
  \setbox0=\vbox{\singlespace \halign{\tabskip 0pt \strut ##\hfil\cr            
   \noalign{\global\dt@ptrue}#1\crcr}}                                          
  \line{\hskip 0.5\hsize minus 0.5\hsize \box0\hfil} \medskip }                 

\def\endletter{\ifnum\pagenumber=1 \vskip\letterbottomfil\supereject            
\else \vfil\supereject \fi}                                                     
\newbox\letterb@x                                                               
\def\lettertext{\par\unvcopy\letterb@x\par}                                     
\def\multiletter{\setbox\letterb@x=\vbox\bgroup                                 
      \everypar{\vrule height 1\baselineskip depth 0pt width 0pt }              
      \singlespace \topskip=\baselineskip }                                     
\def\letterend{\par\egroup}                                                     
%
%
%
\newskip\frontpageskip                                                          
\newtoks\pubtype                                                                
\newtoks\Pubnum                                                                 
\newtoks\pubnum                                                                 
\newif\ifp@bblock  \p@bblocktrue                                                
\def\PH@SR@V{\doubl@true \baselineskip=24.1pt plus 0.2pt minus 0.1pt            
             \parskip= 3pt plus 2pt minus 1pt }                                 
\def\PHYSREV{\paperstyle\PhysRevtrue\PH@SR@V}                                   
\def\titlepage{\FRONTPAGE\paperstyle\ifPhysRev\PH@SR@V\fi                       
   \ifp@bblock\p@bblock\fi}                                                     
\def\nopubblock{\p@bblockfalse}                                                 
\def\endpage{\vfil\break}                                                       
\frontpageskip=1\medskipamount plus .5fil                                       
\pubtype={\tensl Preliminary Version}                                           
\Pubnum={$\caps NUB - \the\pubnum $}                                     
\pubnum={0000}                                                                  
\def\p@bblock{\begingroup \tabskip=\hsize minus \hsize                          
   \baselineskip=1.5\ht\strutbox \topspace-2\baselineskip                       
   \halign to\hsize{\strut ##\hfil\tabskip=0pt\crcr                             
   \the\Pubnum\cr \the\date\cr \the\pubtype\cr}\endgroup}                       
\def\title#1{\vskip\frontpageskip \titlestyle{#1} \vskip\headskip }             
\def\author#1{\vskip\frontpageskip\titlestyle{\twelvecp #1}\nobreak}

\def\address#1{\par\kern 5pt\titlestyle{\twelvepoint\it #1}}                    
\def\andaddress{\par\kern 5pt \centerline{\sl and} \address}                    

\def\abstract{\vskip\frontpageskip\centerline{\fourteenrm ABSTRACT}             
              \vskip\headskip }

%
%
%

\def\\{\relax\ifmmode\backslash\else$\backslash$\fi}                            
\def\globaleqnumbers{\relax\if\equanumber<0\else\global\equanumber=-1\fi}       
\def\nextline{\unskip\nobreak\hskip\parfillskip\break}                          
                       
\def\journal#1&#2(#3){\unskip, \sl #1~\bf #2 \rm (19#3) }                       

\def\topspace{\hrule height 0pt depth 0pt \vskip}                               

\let\int=\intop                                                
\def\prop{\mathrel{{\mathchoice{\pr@p\scriptstyle}{\pr@p\scriptstyle}{          
                \pr@p\scriptscriptstyle}{\pr@p\scriptscriptstyle} }}}           
\def\pr@p#1{\setbox0=\hbox{$\cal #1 \char'103$}                                 
   \hbox{$\cal #1 \char'117$\kern-.4\wd0\box0}}                                 
\def\lsim{\mathrel{\mathpalette\@versim<}}                                      
\def\gsim{\mathrel{\mathpalette\@versim>}}                                      
\def\@versim#1#2{\lower0.2ex\vbox{\baselineskip\z@skip\lineskip\z@skip          
  \lineskiplimit\z@\ialign{$\m@th#1\hfil##\hfil$\crcr#2\crcr\sim\crcr}}}        
%
%
%
\let\sec@nt=\sec                                                                
\def\sec{\relax\ifmmode\let\n@xt=\sec@nt\else\let\n@xt\section\fi\n@xt}         
\def\obsolete#1{\message{Macro \string #1 is obsolete.}}                        
\def\firstsec#1{\obsolete\firstsec \section{#1}}                                
\def\firstsubsec#1{\obsolete\firstsubsec \subsection{#1}}                       
\def\thispage#1{\obsolete\thispage \global\pagenumber=#1\frontpagefalse}        
\def\thischapter#1{\obsolete\thischapter \global\chapternumber=#1}              
\def\nextequation#1{\obsolete\nextequation \global\equanumber=#1                
   \ifnum\the\equanumber>0 \global\advance\equanumber by 1 \fi}                 
\def\BOXITEM{\afterassigment\B@XITEM\setbox0=}                                  
\def\B@XITEM{\par\hangindent\wd0 \noindent\box0 }                               
%
%
\catcode`@=12 
%
%

\hoffset=0.5in
\voffset=0.5in
\baselineskip=32pt
\overfullrule=0pt
\centerline{\bf Kinetic Friction due to Ohm's Law Heating}
\centerline{J. B. Sokoloff, Physics Department and}
\centerline{Center for Interdisciplinary Research on complex Systems,}
\centerline{Northeastern University, Boston, Massachusetts 02115}
\endpage
\centerline{\bf Kinetic Friction due to Ohm's Law Heating}
\centerline{J. B. Sokoloff, Physics Department and}
\centerline{Center for Interdisciplinary Research on complex Systems,}
\centerline{Northeastern University, Boston, Massachusetts 02115}
\medskip
\centerline{\bf Abstract}

Using both a recent calculation by Bruch of the damping of the motion of a
monolayer
nitrogen film oscillating harmonically on a metallic surface due to Ohm's law
heating and a Thomas-Fermi approximation treatment of the Ohm's law heating
mechanism, which accounts for the nonzero thickness of the
surface region of a metal, it is argued that this mechanism for friction
is able to account for recent measurements of the drop in the friction for a
nitrogen film
sliding over a lead substrate as it goes below its superconducting transition
temperature. Bruch's calculation is also made more transparent by
re-doing the calculation for a film sliding at constant speed, instead of
oscillating. Using this treatment, it is easily shown
that Bruch's calculation is
equivalent to integrating Boyer's solution of the problem
of a charge sliding over a metallic surface
over the charge density of the monolayer nitrogen film.

\noindent {\bf 1. Introduction}

Motivated by attempts[1-3] to explain a recent quartz crystal microbalance 
experiment[4] which shows a rapid drop in the friction of a film of nitrogen 
molecules sliding on a lead substrate, on dropping below the 
superconducting transition temperature $T_c$ of the substrate, Bruch has 
recently done a calculation of the electronic contribution to the 
friction for a monolayer nitrogen film 
executing simple harmonic motion on a metallic substrate[5]. 
In contrast to calculations of a single molecule moving  
on the substrate[6,8], which require that the molecule possess a larger dipole 
moment or charge than is generally accepted for adsorbed molecules, 
in order to explain the experimental results of Ref. 4 [1-3], 
Bruch's results suggest that the field due to the quadrupole moment 
of the nitrogen molecule can explain the microbalance experiments[5] if 
the molecules form a monolayer film. The reason for this is that for a 
monolayer film the field 
inside the metallic substrate on which the film is moving falls exponentially 
to zero over a distance of the order of a lattice constant of the film 
below the surface of the metal. Since the distance over which the field is 
nonzero is much smaller than a mean free path, Bruch pointed out that 
the problem must be treated in the anomalous skin effect regime[7], 
in which only those electrons with velocities 
nearly parallel to the film remain in this region for a sufficient length 
of time to be significantly accelerated by the electric field. 
Consequently, only a fraction $(G\ell)^{-1}$ of the electrons 
(where G is the magnitude of a typical reciprocal lattice vector 
and $\ell$ is the mean free path) can be accelerated by 
the field. This results in an enhancement of the 
effective resistivity of the skin depth region, which leads to an enhancement 
of the rate of dissipation and the contribution to the kinetic friction 
due to Ohm's law heating. 

In addition to the Ohm's law heating mechanism for electronic friction put 
forward in Refs. 1 and 2 as a possible way of explaining the experimental 
results reported in Ref. 4, there is another mechanism for 
electronic friction (actually, the most commonly suggested mechanism for 
this phenomenon) which is due to the creation of electron-hole pairs of 
nonzero energy[9-12]. 
The physical difference between these two mechanisms can be understood 
as follows: In the electron-hole pair mechanism for electronic friction, 
the energy loss due to friction is ascribed to the energy needed to create 
the electron-hole pairs. In contrast, in the Ohm's law heating mechanism 
considered in references 1, 2 and 5 and in this article, the energy loss 
need not be due to the energy cost necessary to create 
electron hole pairs. In this mechanism, if we consider the case in which the 
film is slid along at constant speed (e.g., by an applied force), the electric 
field resulting from the sliding film results in a screening charge near the 
surface of the metal, which is dragged along with the film. This results in 
an electric current. Let us first consider only elastic scattering of the 
electrons (the dominant contribution to the resistivity well below the 
Debye temperature). During the sliding, electrons get scattered elastically 
by impurities 
and other defects. This would result in a reduction in the drift velocity, and 
hence the electric current, except that we force the current to remain 
constant by forcing the film to move at constant speed. In order to 
maintain the current, the electric field acting on the conduction electrons 
due to the film must accelerate them 
in order to maintain the drift velocity. The work done by this 
field is identified with the contribution to the dissipation produced by 
Ohm's law heating. In the Ohm's law heating mechanism considering 
only elastic scattering of the electrons, the electron-hole 
pairs resulting from the scattering of the electrons by defects in the 
substrate have zero energy because the scattering is elastic. In addition 
to the contribution due to elastic scattering, there is also inelastic 
scattering of the electrons by phonons, which results in an additional energy 
loss from energy transferred from the electrons to the phonons. Below $T_c$, 
the above scattering mechanisms do not occur for the superconducting 
electrons because of the gap in their excitation spectrum. The 
screening charge will be transported entirely by the superconducting electrons 
because they can flow without electrical resistance, and hence are able 
to short circuit the current due to the normal electrons. Since the 
superconducting electrons are not scattered, the Ohm's law heating mechanism 
for dissipation (and hence kinetic friction) described above does not operate. 

Persson has argued that the electron-hole pair mechanism should dominate over 
the Ohm's law heating mechanism by three orders of magnitude for a charged 
ion moving above the surface of a metallic 
substrate[13]. Since this mechanism depends on the density of normal electrons, 
which does not drop rapidly on falling below the superconducting transition 
temperature, however, it cannot account for the experimental result 
reported in Ref. 4.
The calculation of the Ohm's law heating contribution of the friction, 
including the anomalous skin effect 
conductivity, as suggested by Bruch[5], however, can be of the same order of 
magnitude as the electron-hole contribution to the friction or greater. Since 
the 
resistivity drops to zero over a relatively small temperature range on dropping
below $T_c$, the latter mechanism for electronic friction drops rapidly in 
much the same way as in Ref. 4, and, as we shall see, it is large enough to 
explain the experimental observations.

In Bruch's treatment, the simple harmonic motion 
of the film results in an electric field with many harmonics of the 
frequency of the film's oscillations, which is an unnecessary complication 
in his treatment. In contrast, if one considers the 
film to be sliding over the substrate at constant speed, the field will possess 
only a single harmonic of the "washboard frequency," which is the 
characteristic frequency in this formulation of the problem. 
In section II, Bruch's calculation is reformulated by 
assuming that the film slides at constant velocity, rather than executing 
simple harmonic motion. It is expected  
that this will give the same value for the friction, and it 
simplifies the calculation. Furthermore, doing the calculation in 
this way makes it possible to verify that  
the force of friction acting on the sliding film calculated from 
the Ohmic heating inside the metallic substrate is equal to the 
friction found by calculating the force on the film due to its image field 
acting back on it, as is required by energy conservation. 
(This is not possible in Bruch's treatment because 
when averaged over the oscillation period of the film, the force of friction 
averages to zero.) It is argued in section III, that this 
version of Bruch's treatment 
is identical to Boyer's treatment[6], which is identical to the treatment 
of the problem due to Tomassone and Widom[8], used in Refs. 1 and 2. 
Since Ref. 13 stresses the necessity of taking into account the nonzero 
thickness of the surface region, and since, Bruch's  work is a 
classical treatment of the surface of a metal (i.e., one which assumes the 
surface region has zero thickness), 
in section IV, a calculation is presented which takes into 
account the fact that the surface region, when treated quantum mechanically, 
has a nonzero thickness, in contrast to the zero thickness that it has in 
treatments of this problem using classical electrodynamics [1-3,5-8].

\noindent {\bf 2. Bruch's Idea Applied to a Uniformly Sliding Film} 

Let us take the region with $z<0$ to be occupied by the metal, with the region 
with $z>0$ occupied by free space. Bruch [5] writes the z-component of the 
electric field in terms of its Fourier transform on the x and y coordinates, 
$E({\bf G},z)$,
$$E_z ({\bf r})=\sum_{\bf G} E({\bf G},z)e^{i{\bf G}\cdot {\bf r}}, \eqno (1)$$
where ${\bf G}$ denotes a reciprocal lattice vector of the nitrogen 
film. Using the requirement that the field have zero divergence (in regions 
with zero charge density), the field components parallel to the surface are 
given by
$${\bf E_{||}}({\bf r})=\sum_{\bf G} (i{\bf G}/G^2)
\partial E({\bf G},z)/\partial z e^{i{\bf G}\cdot {\bf r}}. \eqno (2)$$
In regions in which there is no net charge density the 
current density ${\bf J} ({\bf r})$ can similarly be expressed 
in terms of the Fourier transform of its z-component. 
In Bruch's work, the film is assumed to execute simple harmonic motion of 
frequency $\Omega$ as a rigid unit, which means that the time dependent 
fields are obtained by replacing ${\bf r}$ by ${\bf r}-{\bf A}cos(\Omega t)$, 
where the vector ${\bf A}$ has magnitude equal to the amplitude and direction 
in the direction of motion of the film. 
When this substitution is made in Eqs. (1) and (2), the time dependence of the 
fields is a sum of harmonics $K\Omega$, where K is an integer, with Bessel 
function coefficients $J_K ({\bf G}\cdot {\bf A})$. In Bruch's solution it 
is necessary to deal with all of the time Fourier components. In this 
section, I 
propose that the inverse slip-time, which Bruch obtains, can be obtained 
much more easily by considering the damping of a film moving at constant 
velocity {\bf v} instead. For the present treatment, in which the film 
slides at constant velocity ${\bf v}$, we obtain the time dependence simply by 
replacing {\bf r} by {\bf r}-{\bf v}t, which results in a 
field which contain only a single time Fourier component for each 
reciprocal lattice vector, with frequency equal to the "washboard frequency,"
${\bf G}\cdot {\bf v}$, and there is no sum over harmonic with Bessel function 
coefficients. Because of the linear nature of Ohm's law, this method must give 
the same value for the slip-time as Bruch obtains. 
Bruch obtains a relationship between the 
the time and space Fourier transforms of the z-component of the current 
density inside the film, and the z-component of the time and space Fourier 
component of the field on the surface of the metal, just inside the metal 
[Eq. (2.10) in Bruch's paper] by solving the linearized Boltzmann equation 
simultaneously with Faraday's and Ampere's laws. Because the frequencies 
involved in this problem are quite small, Bruch solves these equations 
in the zero frequency limit. As a consequence, each time Fourier component 
of the field and the current density satisfies the same equations (since 
in the zero frequency limit, the coefficients multiplying the fields in the 
equations are obviously independent of frequency). Thus, Bruch's 
solution of Boltzmann's equation with Faraday's and Ampere's law can be 
equally well applied to the present case of a film sliding at constant 
velocity,  for which there is only one time Fourier component for each 
reciprocal lattice vector ${\bf G}$, e.g., ${\bf E} ({\bf G},z)$ for the 
z-component of the electric field. One obtains for the relationship between 
the Fourier transforms of the z-component of the current density and field 
just below the surface of the metal
$$J({\bf G},z=0^-)=\sigma_{\bf G} E(({\bf G},z=0^-), \eqno (3)$$
where $\sigma_{\bf G}=3\sigma (1-p)/(4G\ell)$, where $\ell$ is the mean free 
path,  p is the fraction of the conduction electrons which are specularly 
reflected at the surface of the metal at z=0 and $\sigma$ is the Ohm's 
law electrical conductivity. This is Bruch's Eq. (2.10). 
Substituting Eq, (3) in the standard boundary condition [14]
$$J_z({\bf r},z=0^-)=
(4\pi)^{-1}{\partial\over\partial t}[E_z ({\bf r},z=0^-)-E_z ({\bf r},z=0^+)],
\eqno (4)$$
we obtain
$$\sum_{\bf G} \sigma_{\bf G}E({\bf G},z=0^-)e^{i{\bf G}\cdot {\bf r}}=$$
$$(4\pi)^{-1}\partial [E_z (z=0^-)-E_z (z=0^+)]/\partial t=$$
$$i(4\pi)^{-1}\sum_{\bf G} ({\bf G}\cdot {\bf v})[E({\bf G},z=0^-)-
E({\bf G},z=0^+)]e^{i{\bf G}\cdot {\bf r}}, \eqno (5)$$
where we have used the fact that E and J have the time dependence 
$e^{-i{\bf G}\cdot {\bf v}t}$ for a film sliding at constant speed, for the 
reasons given above.  Eq. (5) can be written as 
$$(1+i\lambda_{\bf G})E({\bf G},0^-)=E({\bf G},0^+)=
B_{\bf G}e^{Gz}+E_i({\bf G},0^+),
\eqno (6)$$
where $E_i({\bf G},z)$ and $E({\bf G},z)$ are the Fourier transform of the z component of the 
contribution to the field from the film in the absence of the substrate 
and inside the metal, respectively, and 
$B_{\bf G}e^{Gz}$ is the contribution of the Fourier transform of the z 
component of 
the field outside of the substrate due to the charge density induced by  the 
film. $B_{\bf G}$ is a constant to be determined by the boundary conditions. 
The parameter 
$\lambda_{\bf G}=4\pi\sigma_{\bf G}/({\bf G}\cdot {\bf v})$, where 
$({\bf G}\cdot {\bf v})$ is the "washboard" frequency of the film. 
In addition to Eq. (6), we must require continuity of the component of the 
field parallel to the surface, which is given by 
$$dE({\bf G},z)/dz|_{z=0^-}=[d(B_{\bf G}e^{Gz})/dz+
dE_i({\bf G},z)/dz]|_{z=0^+}, 
\eqno (7)$$
which using the result
$$E({\bf G},z)=E({\bf G},z=0^-)e^{Gz}$$
given in Eq. (A13) of the appendix of Ref. 5 (It follows from 
the solution of Ampere's and Faraday's law in the zero frequency limit.) gives 
$$-E({\bf G},0^+)=B_{\bf G}-E_i ({\bf G},0^-). \eqno (8)$$
The solution to Eqs. (6) and (8) is
$$E({\bf G},0^+)=[2/(2+i\lambda_{\bf G})]E_i({\bf G},0^-), \eqno (9a)$$
and
$$B_{\bf G}=[i\lambda_{\bf G}/(2+i\lambda_{\bf G})]E_i({\bf G},0^-)\approx 
(1+2i/\lambda_{\bf G}+...)E_i({\bf G},0^-), \eqno (9b)$$ 
where $E_i({\bf G},Z)$ can be found using 
$$E_i ({\bf r})=\int d^3 r'\rho ({\bf r}) {{\bf r}-{\bf r'}\over 
|{\bf r}-{\bf r'}|^3}, \eqno (10)$$
where $\rho ({\bf r})$ 
is the film's charge density. It can be crudely modeled 
by three charges along the axis of a molecule ${\bf\ell}_{\beta}$, in the 
small $\ell_{\beta}$ limit, two
charges of charge +q at the outer edges of the molecule and a charge -2q at
its center. The value of q is chosen such that the resulting 
quadrupole moment has the experimental value $\theta$. Then, in the unit cell 
near the origin 
$$\rho ({\bf r})=q\delta(z+h)\sum_{\beta=1,2}
[\delta^{(2)} ({\bf r}-{\bf\rho}_{\beta}-{\bf\ell}_{\beta})+
\delta^{(2)}  ({\bf r}-{\bf\rho}_{\beta}+{\bf\ell}_{\beta})$$
$$-2\delta^{(2)}  ({\bf r}-{\bf\rho}_{\beta})], \eqno (11)$$
where $\delta^{(2)}$ denotes a two dimensional delta function and
${\bf\rho}_{\beta}$ denotes the position of a molecule in the unit cell. 
Writing  
$${{\bf r}-{\bf r'}\over |{\bf r}-{\bf r'}|^3}=
4\pi (2\pi)^{-3}\int d^3 k (i {\bf k}/k^2)
e^{i{\bf k}\cdot ({\bf r}-{\bf r'})}$$
and substituting in Eq. (10), we obtain 
$$E_i({\bf G},0^-)=-(\pi\theta/A_c)e^{-Gh}
\sum_{\beta}e^{-i{\bf G}\cdot {\bf \rho}_{\beta}}
({\bf G}\cdot\hat\ell_{\beta})^2, \eqno (12)$$
where ${\bf\rho}_{\beta}$ is the location of the $\beta^{th}$ molecule in the 
nitrogen film unit cell, $\hat\ell_{\beta}$ is the symmetry axis of 
$\beta^{th}$ molecule, and $\theta$ is the quadrupole moment of 
a single molecule. The inverse slip-time found by calculating the force exerted on the 
film by the image field outside 
the metal, 
$$\sum_{\bf G}(i{\bf G}/G^2)
\partial [B_{\bf G}e^{Gz}e^{i{\bf G}\cdot {\bf r}}]/\partial z,$$
is given by $(NMv)^{-1}$ times the force or
$$(NMv)^{-1}\int d^3 r \rho ({\bf r}) \sum_{\bf G}(i{\bf G}/G^2) 
\partial [Be^{Gz}e^{i{\bf G}\cdot {\bf r}}]/\partial z. 
\eqno (13)$$
From Eqs. (9b) and (13), we obtain for the inverse slip-time
$$\tau^{-1}=({\theta^2\over 3(1-p)\sigma A_c Mv})|\sum_{\bf G} 
e^{-2Gh} (4G\ell)({\bf G}\cdot {\bf v})
({\bf G}/G)$$
$$|\sum_{\beta} ({\bf G}\cdot\hat\ell_{\beta})^2 
e^{i{\bf G}\cdot {\bf\rho}_{\beta}}|^2|, \eqno (14)$$
which gives 
$$\tau^{-1}=({\theta^2\over 3(1-p)\sigma A_c M})\sum_{\bf G}(4G\ell)(G_x^2/G)
e^{-2Gh}$$
$$|\sum_{\beta} ({\bf G}\cdot\hat\ell_{\beta})^2
e^{i{\bf G}\cdot {\bf\rho}_{\beta}}|^2, \eqno (15)$$
where we have taken v to be along the x-axis. 
By energy conservation, this expression for $\tau^{-1}$ must be equal to the 
value calculated by setting the power loss due to Ohm's law heating equal 
to $NMv^2/\tau$, the rate at which the viscous electronic contribution to the 
friction force, $NMv/\tau$, does work on the film. Using the method used in this 
paper of assuming that the film is sliding at a speed v, rather than 
oscillating (as assumed in Bruch's paper), the Ohm's law heating power is 
given by 
$$P=2NA_c \sum_{\bf G}G^{-1}|E_i ({\bf G},0^-)|^2 
({\bf G}\cdot {\bf v})/\sigma_{\bf G} \eqno (16)$$
The factor $NA_c$ appears 
because when P is calculated by integrating ${\bf J}({\bf r})\cdot {\bf E} 
({\bf r})$ over the volume of the metal, the volume integral involves 
evaluating the integral $\int d^2 r e^{i({\bf G}-{\bf G'})\cdot {\bf r}}=
NA_c\delta_{{\bf G},{\bf G'}}.$ 
We obtain with this proceedure[15], substituting $E_i$ from Eq. (12) in Eq. (16), 
the value for $\tau^{-1}$ obtained in Eq. (15). 
 
\noindent {\bf 3. Treatment of Electronic Friction using Boyer's Solution}

Boyer[6] solves the problem of a charge or electric dipole moving above (i.e., 
outside the metal) and parallel to the surface of the metal. He solves the 
electrodynamics problem subject to the same boundary conditions as Bruch uses 
[14], [Eq. (4) above], which for Boyer's problem 
is written as 
$$J_z=\sigma E=(4\pi)^{-1}\partial E_z/\partial t=(4\pi)^{-1} {\bf v}\cdot 
\nabla E_z. \eqno (17)$$
When terms of up to first order in v are kept, we obtain Boyer's result 
for the electric field. The force exerted by this field on the moving 
charge gives the force of friction due to Ohm's law heating in the metal. 
The force of friction found by the formalism due to Tomassone and Widom[8] 
gives the same friction and hence is believed to be equivalent to Boyer's 
calculation. 

Boyer[6] finds, in addition to the electrostatic field, a contribution to the 
electric field above the substrate linear 
in the velocity v of a point charge q sliding above the substrate, given by
$${\bf E}({\bf r})=-(qv/2\pi\sigma) {\partial\over\partial x}[{{\bf r}-{\bf r'}
\over |{\bf r}-{\bf r'}|^3}], \eqno (18)$$
assuming that the sliding velocity is in the x-direction, 
where ${\bf r'}=vt\hat x+h\hat z$ is the location of the 
moving charge (where $\hat x$ and $\hat z$ are unit vectors in the x and z 
directions, respectively). In order to apply this result to 
a monolayer film of charge density $\rho ({\bf r})$, let us multiply Eq. (18) 
by  $\rho ({\bf r})$ and integrate over volume to obtain ${\bf E}({\bf r}),$
$${\bf E}({\bf r})=(v/2\pi\sigma)\int d^3 r'\rho ({\bf r'})(4\pi)$$
$$\times (2\pi)^{-3}\int d^3 k({\bf k}k_x/k^2)
e^{i{\bf k}\cdot ({\bf r}-{\bf r'})} \eqno (19)$$
where we have written the field in Eq. (19) in terms of its Fourier transform. 
For a periodic 
monolayer film a height h above the surface of the substrate, $\rho ({\bf r})$ 
has the form
$$\rho ({\bf r})=\delta (z-h)\sum_{\bf G}\rho_{\bf G}
e^{i{\bf G}\cdot({\bf r}-vt\hat x)},
\eqno (20)$$
where ${\bf G}$ denote the reciprocal lattice vectors of the film and 
$\rho_{\bf G}=A_c^{-1}\int_u d^2 r \rho_2 ({\bf r})
e^{-i{\bf G}\cdot {\bf r}}$, where the 
u on the integral sign signifies an integral over a unit cell of the film,
 $A_c$ is the unit cell area 
and $\rho_2 ({\bf r})$ is the charge per unit area of the film. For simplicity, 
we are modeling the film by a collection of point charges. If the nonzero 
size of the charges in the film were taken into account, there would 
be form factors introduced in the summations over ${\bf G}$, which would 
fall off rapidly with increasing magnitude of ${\bf G}$. This can be 
approximately accounted for by including only the first one or two terms in 
the sums.
Substituting Eq. (20) in Eq. (19) gives
$${\bf E}=2v(2\pi\sigma)^{-1} \sum_{\bf G}\rho_{\bf G} 
\int dk_z ({{\bf k} G_x\over k_z^2+G^2})
e^{i{\bf k}\cdot [{\bf r_{||}}-vt\hat x-(z+h)\hat z]}, \eqno (21)$$
where ${\bf r_{||}}$ is the projection of ${\bf r}$ in the plane of the film, 
and where the x and y components of ${\bf k}$ are equal to the x and y 
components of film reciprocal lattice vectors. 
For components of ${\bf E}$ parallel to 
the substrate we obtain on performing the integral over $k_z$ 
$${\bf E}({\bf r})=(v/\sigma)
\sum_{\bf G}\rho_{\bf G}({\bf G}G_x/G)e^{-G(z+h)}
e^{i({\bf G}\cdot {\bf r}+G_x vt)}. \eqno (22)$$
To find $\rho_{\bf G}$, we model the charge density of of each molecule 
by three charges as was done in the last section. 
Then, substituting $\rho ({\bf r})$ from Eq. (11) in the integral for 
$\rho_{\bf G}$ under Eq. (20), we obtain
$$\rho_{\bf G}=-4qA_c^{-1}\sum_{\beta}
sin^2({\bf G}\cdot {\bf\ell}_{\beta}/2)
e^{i{\bf G}\cdot {\bf\rho}_{\beta}}. \eqno (23)$$
Therefore,
$${\bf E}({\bf r})=-v 4qA_c^{-1}\sigma^{-1}\sum_{\beta,{\bf G}}
sin^2({\bf G}\cdot {\bf\ell}/2)$$
$$\times ({\bf G}G_x/G)e^{-G(z+h)}
e^{i[{\bf G}\cdot ({\bf\rho}_{\beta}+{\bf r})-G_x vt]}, \eqno (24)$$
which reduces to an expression resembling Bruch's for the field in the small 
${\bf \ell_{\beta}}$ limit with the quadrupole moment of the molecule 
$\theta$ equal to $q\ell_{\beta}^2$, if we pretend that the system is not 
in the anomalous skin effect regime, and replace $\sigma_{\bf G}$ by 
$\sigma$ for the purpose of making a comparison. The force of friction 
acting on the film is given by
$$F=(1/2)Re\int d^3 r\rho^* ({\bf r}){\bf E} ({\bf r})=
-(N\theta^2 {\bf v}/\sigma)$$
$$\times A_c^{-1}\sum_{{\bf G}}| \sum_{\beta} 
({\bf G}\cdot\hat\ell_{\beta})^2 e^{i{\bf G}\cdot {\bf\rho}_{\beta}}|^2
({\bf G}G_x/G), \eqno (25)$$
where N is the number of molecules in the film, 
and thus the inverse slip-time $\tau^{-1}$ is given by
$$\tau^{-1}=(F/NMv)=$$
$$A_c^{-1} (\theta^2/M\sigma)\sum_{{\bf G}}
 | \sum_{\beta}({\bf G}\cdot\hat\ell_{\beta})^2 
e^{i{\bf G}\cdot {\bf\rho}_{\beta}} | ^2
(G_x^2/G)e^{-2Gh}, \eqno (26)$$
where M is the mass of an adsorbed molecule and $\hat\ell_{\beta}$ is a unit 
vector along ${\bf\ell}_{\beta}$. If as discussed above, we 
replace  $\sigma_{\bf G}$ of section II by $\sigma$, 
Eq. (15) becomes identical to 
Eq. (26). This demonstrates the equivalence of Bruch's and Boyer's 
treatments. 

\noindent {\bf 4. A Treatment of the Problem which Includes the Nonzero Width of the 
Surface Region}

Both Bruch's treatment of the problem and Boyer's are 
based on the classical model for a metallic surface, in which the electronic 
charge density drops to zero immediately on leaving the metal. In a 
quantum mechanical treatment, in contrast, the electronic charge density 
drops to zero over a distance of the order of a couple of  Angstroms. 
The film almost 
certainly resides in a region just outside the bulk of the metal, in which 
the electronic charge density is nonzero (although it is decaying exponentially 
here). Ying, et. al., studied electron screening in the 
surface region using the Thomas-Fermi approximation[16]. They find that a charge 
placed at the surface will be screened, with a screening length not much 
longer than that in the bulk metal. The bulk metal Thomas Fermi expression 
for the screening charge density of a point charge can be used to model the 
charge density in the surface region analytically if one fits the 
screening length to that found in Ref. 16. The bulk Thomas 
expression for the screening charge density for a point charge is 
$$\rho_s ({\bf r})=-(4\pi)^{-1}qk_s^2 e^{-k_s r}/r=$$
$$-qk_s^2(2\pi)^{-3}\int d^3 k {1\over k^2+k_s^2} e^{i{\bf k}\cdot {\bf r}}. 
\eqno (27)$$
where q is the point charge whose screening is being considered and $k_s$ is 
the inverse Thomas Fermi screening length, which we will take 
here to be a parameter to be fit to Ying, et. al.'s calculations[16]. 
In Ref. 16, the quantity
$$\int dx dy\rho ({\bf r}) \eqno (28)$$
is plotted as a function of z. The model of Eq. (27) gives for this quantity
$$-(1/2) q k_s e^{-k_s |z|}, \eqno (29)$$
which strongly resembles the quantity plotted in Fig. 1 of Ref. 16. Then to fit 
the present approximate model to the results of Ref. 16, we can simply choose 
a value of $k_s$ for which Eq. (26) reproduces each of the plots in Fig. 1 of 
that reference. This method is rigorous when one can use the quasi-classical 
approximation, which is accurate if typical values of the screening length 
are much smaller than the thickness of the surface region. (See the appendix.) 
When this is not an appropriate limit, for example if the film is further 
out from the bulk of the metal, there is asymmetry in the screening 
charge density[17], 
but it is not such an extreme asymmetry as to invalidate using a spherically 
symmetric screening charge density, as is done here as a first approximation. 

On the basis of this model, we conclude that when a film with charge density 
$\rho ({\bf r})$ moves along the surface of the metal with a velocity 
${\bf v}$, the screening charge density given by
$$\rho_s ({\bf r})=-(2\pi)^{-3} k_s^2 \int d^3 r'\int d^3 k 
{e^{i{\bf k}\cdot ({\bf r}-{\bf r'})}\over k^2+k_s^2}\rho ({\bf r'}) \eqno (30)
$$
moves with the same velocity resulting in a current density ${\bf J_s}=
{\bf v}\rho_s ({\bf r})$. Substituting for the $\rho ({\bf r'})$ 
in terms of its Fourier transform, we obtain
$${\bf J_s} ({\bf r})=k_s^2 (2\pi)^{-1}\sum_{\bf G}\int d k_z
{e^{i{\bf G}\cdot {\bf r}} e^{ik_z (z-h)}\over G^2+k_s^2+k_z^2}
\rho_{\bf G}=$$
$$(1/2)k_s^2 {\bf v}A_c^{-1}\sum_{\bf G}e^{i{\bf G}\cdot {\bf r}}
{exp[-(G^2+k_s^2)^{1/2}|z-h|]\over (G^2+k_s^2)^{1/2}} \rho_{\bf G}. 
\eqno (31)$$
The resulting Ohmic heating contribution to the force of friction can be 
found from
$$F_{fric}v=\sigma^{-1}\int |{\bf J_s}|^2 d^3 r. \eqno (32)$$
Substituting Eq. (31) into Eq. (32) while replacing $\sigma$ by an 
effective conductivity $\sigma_{\bf G}\approx\sigma/G\ell$ to account for 
the anomalous skin effect [7] (like the 
effective conductivity used in section II) and 
placing it within the summation over 
${\bf G}$, we obtain for the force of 
friction 
$${\bf F_{fric}}=A k_s^4 {\bf v}\sum_{\bf G} \sigma_{\bf G}^{-1}
(G^2+k_s^2)^{-3/2}|\rho_{\bf G}|^2. \eqno (33)$$
Then, we have for the inverse slip-time
$$\tau^{-1}={F_{fric}\over NMv}=$$
$${k_s^4\theta^2 \over 2A_c M}\sum_{\bf G}\sigma_{\bf G}^{-1}
(G^2+k_s^2)^{-3/2}|\sum_{\beta}
({\bf G}\cdot\hat\ell_{\beta})^2 e^{i{\bf G}\cdot{\bf \rho_{\beta}}}|^2, 
\eqno (34)$$
where we have substituted for $\rho_{\bf G}$ using Eq. (23), where A 
is the area of the film $A_c$ is the area of a unit cell and N is the 
number of unit cells in the film. 
For $k_s\approx 10^8 cm^{-1}$ and $\theta\approx 10^{-26}esu$, we obtain
$\tau^{-1}\approx 10^{11} s^{-1},$ which is much larger than the electron-hole 
creation mechanism discussed earlier[13]. The quantity $\tau^{-1}$ found from 
Eq. (34) is about four orders of magnitude larger than $\tau^{-1}$ found 
from Eq. (15) because Eq. (34) does not contain the factor $e^{-2Gh}$, which 
appears in Eq. (22) (where $G=2\AA^{-1}$ and $h=2\AA$).
If we use a value for $k_s$ 
a factor of 5 smaller (which is not unreasonable considering that we are 
in the surface region where $k_s$ could be noticeably smaller than its 
bulk value because the conduction electron density is smaller here), 
$\tau^{-1}$ becomes of order $10^8 s^{-1}$, comparable to the experimental 
value[4]. 

Bruch's use of a classical (i.e., zero thickness) surface would be valid 
if the film resided well above the surface where the (classical) method 
of electrical images should be valid[17]. The treatment in this section 
assumes that the film resides in a part of the surface region in 
which the electron density is closer to its bulk value. The two treatments 
bracket the true situation, in which the film lies between these two extremes. 
Since both treatments give a large enough magnitude for $\tau^{-1}$ to 
account for the experimental results, one can say with confidence that 
the Ohm's law heating contribution to the friction is of sufficiently large 
magnitude to account for the experimental results of Ref. 4, as put 
forward in Refs. 1 and 2.

\noindent {\bf 5. Conclusions}

Bruch's calculation of the electronic friction acting on a film of nitrogen 
molecules harmonically oscillating on a metallic substrate, is re-done 
for the simpler case in which the film is sliding at constant speed. This 
treatment, which should give the same value for the slip-time as Bruch's 
treatment, clarifies Bruch's treatment and allows one to easily demonstrate 
that the force of friction 
found by calculating the force of the image charge acting back on the 
film and by calculating the Ohm's law heating inside the metallic substrate 
are equal. It also allows one to demonstrate the equivalence of Bruch's 
treatment with that due to Boyer of a charge sliding over a metallic surface. 
Since Boyer's treatment is equivalent to that used in Refs. 1 and 2, one 
is confident in saying that Bruch's treatment of the problem is an extension 
of these methods, which allows one to include the anomalous skin effect.

Persson has argued that the contribution to the friction due to the creation 
of electron-hole pairs should dominate over the Ohm's law heating 
mechanism by three orders of magnitude for a charged ion moving above the 
surface of a metallic 
substrate[13]. Since this mechanism depends on the density of normal electrons, 
which does not drop rapidly on falling below the superconducting transition 
temperature, it cannot account for the experimental result reported in Ref. 4.
The calculation of the Ohm's law heating contribution of the friction 
presented in the previous sections, including the anomalous skin effect 
conductivity, as suggested by Bruch[5], however, can be of the same order of 
magnitude as the electron-hole contribution to the friction. Since the 
resistivity drops to zero over a relatively small temperature range on dropping
below $T_c$, the latter mechanism for electronic friction drops rapidly in 
much the same way as in Ref. 4, and, as we have seen, it is large enough to 
explain the experimental observations.

\noindent {\bf acknowledgments}

I wish to thank the Department of Energy for their financial support 
(Grant DE-FG02-96ER45585). I also 
wish to thank L. W. Bruch for discussions that I have had with him about his 
work on this problem, as well as R. Markiewicz and A. Widom.

\noindent {\bf Appendix: Quasiclassical Treatment of the Screening of 
a Charge in the Surface Region}

In the surface region, where the electron charge density is decreasing from its 
bulk value down to zero, the wave functions in the jelleum model must take the 
form 
$$\psi_{\bf k} ({\bf r})=e^{i{\bf k}\cdot {\bf r}} f_{\bf k} (z), \eqno (A1)$$
where $f_{\bf k} (z)$ drops from the value it has for z in the bulk region 
to zero for z well above the surface.  
Then the electron charge density is equal to
$$\sum_{\bf k} n_{\bf k}|\psi_{\bf k} ({\bf r})|^2=
\sum_{\bf k} n_{\bf k}|f_{\bf k} (z)|^2, \eqno (A2)$$
where $n_{\bf k}$ is the Fermi function, 
$(e^{(\epsilon ({\bf k})-\mu)/k_B T}+1)^{-1}$, where $\epsilon ({\bf k})$ is 
the electron energy of wavevector ${\bf k}$ and $\mu$ is the chemical 
potential. The potential $\phi ({\bf r})$ satisfies Poisson's equation
$$\nabla^2\phi ({\bf r})=-4\pi \sum_{\bf k} n_{\bf k}|f_{\bf k} (z)|^2. 
\eqno (A3)$$
Let us now consider the screening of a point charge q placed at a point
${\bf r}=z_0\hat z$ inside the surface region, where $\hat z$ is a unit vector 
in the z-direction. In the linearized Thomas-Fermi treatment of screening, we 
assume that $\phi ({\bf r})$ is changed by a small amount 
$\delta\phi ({\bf r})$ because of the point charge, add this change in 
the potential to $\mu$ and linearize in $\delta\phi$. Carrying this out we 
find that $\delta\phi ({\bf r})$ satisfies
$$\nabla^2 \delta\phi ({\bf r})= -k_s^2 (z) \delta\phi ({\bf r})+
4\pi\delta({\bf r}), \eqno (A4)$$ 
where 
$$k_s^2 (z)\approx 4\pi\sum_{\bf k}\delta (\epsilon ({\bf k})-\mu) 
|f_{\bf k} (z)|^2 \eqno (A5)$$
in the low temperature limit. Taking the Fourier transform of Eq. (A4) with 
respect to the components of ${\bf r}$ parallel to the surface it becomes 
$$d^2 \delta\phi ({\bf k_{||}},z)/dz^2=
-(k_s^2 (z)-k^2_{||})\delta\phi ({\bf k_{||}},z)+4\pi\delta (z-z_0), \eqno (A6)$$
where $\delta\phi ({\bf k_{||}},z)$ is the Fourier transform on 
$\delta\phi ({\bf r})$ over the coordinates parallel to the surface. Let us 
attempt to find a solution to Eq. (A6) of the form $\delta\phi=e^{-S(z)}$. 
Substituting in Eq. (A6) we obtain
$$[(dS/dz)^2-d^2 S/dz^2-(k_s^2 (z)-k^2_{||})]e^{-S}=4\pi\delta(z-z_0). 
\eqno (A7)$$
If typical values of the screening length are small compared to the width 
of the surface region, we can for most z neglect $d^2 S/dz^2$ 
compared to $(dS/dz)^2$. With this approximation, we obtain a solution 
$$\delta\phi=(4\pi q/2k'_s (z_0)exp[-|\int_{z_0}^{z} k'_s (k')dz'|], 
\eqno (A8)$$
where $(k'_{s})^2=k_{s}^2-k^{2}_{||}$. In the extreme limit in which typical 
$k_s'^{-1}$ values are much smaller than the width of the surface region, we 
can replace the integral in the exponent by $k'_s (z_0) (z-z_0)$ to a 
good approximation. The resulting form for $\delta\phi$ gives an inverse 
Fourier transform on $k_{||}$ proportional to 
$${e^{-k_s (z_0)|{\bf r}-z_0\hat z|}\over |{\bf r}-z_0\hat z|}, \eqno (A9)$$ 
the form of the 
result for Thomas-Fermi of a point charge in the bulk of the metal. 
While the extreme limit of screening length much smaller than the 
width of the surface region is not likely to occur, we do not expect 
the screening in real situations to be so qualitatively different than 
the bulk Thomas Fermi screening. For example, one would not expect the 
screening that occurs parallel to the surface to be qualitatively different 
than that normal to the surface, although there is certainly likely to be 
some anisotropy[17].

\noindent {\bf References}

\noindent
[1] Sokoloff J B, Tomassone M S and Widom A (2000) Phys. Rev. Lett. 
84, 515. \nextline
[2] Novotny T and Velicky B (1999) Phys. Rev. Lett. 83, 4112. \nextline 
[3] Popov V L (1999), Phys. Rev. Lett. 83, 1632. \nextline 
[4] Dayo A, Alnasrallah W  and Krim J (1998) Phys Rev. \nextline Lett. \nextline 80, 1690; 
Renner R L,  Rutledge J E and Taborek P  (1999) Phys Rev Lett 83, 1261; 
Krim J (1999) Phys Rev Lett 83, 1262. \nextline
[5] Bruch L W (2000) Phys Rev B 61, 16201. \nextline
[6] Boyer T H (1974) Phys Rev A 9, 68. \nextline
[7] Pippard A B (1958) in Advances in Electronics and Electron Physics, 
ed L Marton (Academic Press, NY; Mattis D C and Bardeen J (1958), Phys 
Rev 111, 412. \nextline
[8] Tomassone M S  and Widom A (1997) Phys Rev B56, 4938. \nextline
[9] Persson B N J and Andersson S (1984) Phys Rev B 29, 1863. \nextline
[10] Persson B N J and Zaremba E (1984) Phys Rev B 30, 5669. \nextline
[11] Persson B N J and Zaremba E (1985) Phys Rev B 31, 1863. \nextline
[12] Persson B N J and Lang N (1982) Phys Rev B 26, 5409. \nextline 
[13] Persson B N J and Tosatti E (1998) Surf Sci 411, 855; 
Persson B N J (2000) Solid State Communications 115, 145. \nextline
[14] Reitz J R, Milford F J and Christy R W (1993) "Foundations of 
Electromagnetic Theory," 4th edition (Addison-Wesley, Reading Massachusetts), 
section 16-6. \nextline
[15] The Angstrom scale periodic spatial variation of the field of the 
sliding film should not make Ohm's law heating inapplicable We can see this by 
making a Galilean transformation to a reference frame in which the film is 
at rest The Ohm's law heating comes about in this frame from the potential 
from the impurities in the substrate, which in this frame is time dependent 
since the impurities 
are moving, and thus can excite conduction electrons near the Fermi level. 
It is not prevented from doing so by the periodic potential of the film  
(parallel to the surface) because the periodic field cannot 
produce gaps which wipe out the entire Fermi surface. \nextline
[16] Ying S C, Smith J R and Kohn W (1971) J Vac Sci and Tech 9, 575. 
\nextline 
[17] Ying S C, Smith J R and Kohn, W (1973) Phys Rev B 11, 1483. \nextline
\end